\documentclass[runningheads]{llncs}
\usepackage{graphicx}
\usepackage{booktabs} 
\usepackage[normalem]{ulem}
\usepackage{multirow}
\usepackage{bm}
\useunder{\uline}{\ul}{}

\usepackage{enumitem}
\usepackage{color}
\usepackage[ruled]{algorithm2e}
\usepackage{amssymb}
\usepackage{dsfont}
\usepackage{pifont}

\begin{document}
\title{Weakly-Supervised Open-Retrieval Conversational Question Answering}



\author{Chen Qu\inst{1} \and
Liu Yang\inst{1} \and
Cen Chen\inst{2} \and
W. Bruce Croft\inst{1} \and\\
Kalpesh Krishna\inst{1} \and
Mohit Iyyer\inst{1}}

\authorrunning{C. Qu et al.}

\institute{University of Massachusetts Amherst\\
\email{\{chenqu,lyang,croft,kalpesh,m.iyyer\}@cs.umass.edu}
\and
Ant Financial Services Group\\
\email{	chencen.cc@antfin.com}}

\maketitle              
\begin{abstract}
Recent studies on Question Answering (QA) and Conversational QA (ConvQA) emphasize the role of retrieval: a system first retrieves evidence from a large collection and then extracts answers. This open-retrieval ConvQA setting typically assumes that each question is answerable by a single span of text within a particular passage (a span answer). The supervision signal is thus derived from whether or not the system can recover an exact match of this ground-truth answer span from the retrieved passages. This method is referred to as \textit{span-match weak supervision}. However, information-seeking conversations are challenging for this span-match method since long answers, especially freeform answers, are not necessarily strict spans of any passage. Therefore, we introduce a \textit{learned weak supervision} approach that can identify a paraphrased span of the known answer in a passage. 
Our experiments on QuAC and CoQA datasets show that the span-match weak supervisor can only handle conversations with span answers, and has less satisfactory results for freeform answers generated by people. Our method is more flexible as it can handle both span answers and freeform answers.
Moreover, our method can be more powerful when combined with the span-match method which shows it is complementary to the span-match method. We also conduct in-depth analyses to show more insights on open-retrieval ConvQA under a weak supervision setting.

\end{abstract}
%
%
%

\section{Introduction}
\label{sec:intro}
Conversational search and Conversational Question Answering (ConvQA) have become one of the focuses of information retrieval research. 
Previous studies~\cite{quac,coqa} set up the ConvQA problem as to extract an answer for the conversation so far from a \textit{given gold passage}.
Recent work~\cite{orconvqa} has emphasized the fundamental role of retrieval by presenting an Open-Retrieval ConvQA (ORConvQA) setting. This setting requires the system to \textit{learn} to retrieve top relevant passages from a large collection and then extract answers from the passages.

The open-retrieval setting presents challenges to training the QA/ConvQA system. Qu et al.~\cite{orconvqa} adopts a fully-supervised setting, which encourages the model to find the gold passage and extract an answer from it by manually including the gold passage in the retrieval results during training. This \textit{full supervision} setting can be impractical since gold passages may not always be available. In contrast, other studies~\cite{drqa,orqa,Das2019MultistepRI} assume no access to gold passages and identify weak answers in the retrieval results by finding a span that is an exact match to the known answer. We argue that the effectiveness of this \textit{span-match weak supervision} approach is contingent on having only \textit{span answers} that are short, or extractive spans of a retrieved passage. In information-seeking conversations, however, answers can be relatively long and are not necessarily strict spans of any passage. These \textit{freeform answers} can be challenging to handle for span-match weak supervision.

In this work, we introduce a \textit{learned weak supervision} approach that can identify a paraphrased span of the known answer in a retrieved passage as the weak answer. Our method is more flexible than span-match weak supervision since that it can handle both span answers and freeform answers. Moreover, our method is less demanding on the retriever since it can discover weak answers even when the retriever fails to retrieve any passage that contains an exact match of the known answer. By using a weakly-supervised training approach, our ConvQA system can discover answers in passages beyond the gold ones and thus can potentially leverage various knowledge sources. In other words, our learned weak supervision approach makes it possible for an ORConvQA system to be trained on natural conversations that can have long and freeform answers. The choice of the passage collection is no longer a part of the task definition. We can potentially combine different knowledge sources with these conversations since the weak answers can be discovered automatically.

Our learned weak supervisor is based on Transformers~\cite{transformer}. Due to the lack of training data to learn this module, we propose a novel training method for the learned weak supervisor by leveraging a diverse paraphraser~\cite{diverse_paraphrase} to generate the training data. Once the learned weak supervisor is trained, it is frozen and used to facilitate the training of the ORConvQA model. 

We conduct experiments with the QuAC~\cite{quac} and CoQA~\cite{coqa} datasets in an open-retrieval setting. We show that although a span-match weak supervisor can handle conversations with span answers, it is not sufficient for those with freeform answers. For more natural conversations with freeform answers, we demonstrate that our learned weak supervisor can outperform the span-match one, proving the capability of our method in dealing with freeform answers. Moreover, by combining the span-match supervisor and our method, the system has a significant improvement over using any one of the methods alone, indicating these two methods complement each other. 
Finally, we perform in-depth quantitative and qualitative analyses to provide more insight into weakly-supervised ORConvQA. 
Our data and model implementations will be available for research purposes.\footnote{\url{https://github.com/prdwb/ws-orconvqa}}

The rest of our paper is organized as follows. In Section~\ref{sec:relatedwork}, we present related work regarding question answering and conversational question answering. In Section~\ref{sec:our-approach}, we formulate the research question of ORConvQA following previous work and present our weakly-supervised solution. In Section~\ref{sec:exp}, we present our evaluation results on both span answers and freeform answers. Finally, Section~\ref{sec:conclusion} presents the conclusion and future work.
\section{Related Work}
\label{sec:relatedwork}
Our work is closely related to question answering, conversational question answering, session search~\cite{Luo2014WinwinSD,Luo2015LearningTR,Zhou2020RLIRankLT}, and weak supervision and data augmentation~\cite{Li2019InsufficientDC,Chen2020BalancingRL}. We highlight the related works on QA and ConvQA as follows.

\textbf{Question Answering}. 
Most of the previous work formulates question answering either as an answer selection task \cite{wikiqa,trecqa,tanda} or a machine comprehension (MC) task \cite{squad,squad2,GoogleNQ,newsqa}. These settings overlook the fundamental role of retrieval as articulated in the QA task of the TREC-8 Question Answering Track~\cite{trec8}. Another line of research on open-domain question answering addresses this issue by leveraging multiple documents or even the entire collection to answer a question \cite{Marco,TriviaQA,searchqa,quasar,WikiPassageQA}. When a large collection is given as a knowledge source, previous work~\cite{drqa,bertserini} typically uses TF-IDF or BM25 to retrieve a small set of candidate documents before applying a neural reader to extract answers. More recently, neural models are being leveraged to construct learnable rerankers \cite{rank_paragraph,htut_rank,adaptive_doc_retrieval,r3} or learnable retrievers \cite{orqa,Das2019MultistepRI,dpr} to enhance the retrieval performance. Compared to this work on single-turn QA, we focus on a conversational setting as a further step towards conversational search.

\textbf{Conversational Question Answering}.
As an extension of the answer selection and MC tasks in single-turn QA, most research on conversational QA focuses on conversational response ranking~\cite{Yang2018ResponseRW,udc,Yan2016ShallIB,Yan2016LearningTR,Tao2019MultiRepresentationFN,smn,hybrid,iart} and conversational MC~\cite{quac,coqa,hae,ham,flowqa,sdnet,bertflowdelta,graphflow,Qiu2021ReinforcedHB}. A recent paper~\cite{orconvqa} extends conversational QA to an open-retrieval setting, where the system is required to learn to retrieve top relevant passages from a large collection before extracting answers from the passages. Although this research features a learnable retriever to emphasize the role of retrieval in ConvQA, it adopts a fully-supervised setting. This setting requires the model to have access to gold passages during training, and thus is less practical in real-world scenarios. Instead, we propose a learned weakly-supervised training approach that can identify good answers in any retrieved documents. In contrast to the span-match weak supervision~\cite{drqa,orqa,Das2019MultistepRI} used in single-turn QA, our approach is more flexible since it can handle freeform answers that are not necessarily a part of any passage.

\section{Weakly-Supervised ORConvQA}
\label{sec:our-approach}
In this section, we first formally define the task of open-retrieval ConvQA under a weak supervision setting. We then describe an existing ORConvQA model~\cite{orconvqa} and explain how we train it with our learned weak supervision approach.

\subsection{Task Definition}
\label{subsec:task}
We define the ORConvQA task following Qu et al.~\cite{orconvqa}. Given the $k$-th question $q_k$ in a conversation, and all history questions $\{q_i\}_{i=1}^{k-1}$ preceding $q_k$, the task is to predict an answer $a_k$ for $q_k$ using a passage collection $C$. 
Different from Qu et al.~\cite{orconvqa}, we assume no access to gold passages when training the reader. The gold passage for $q_k$ is the passage in $C$ that is known to contain or support $a_k$.

\subsection{An End-to-End ORConvQA System}
\label{subsec:system}
\begin{figure}[t]
    \centering
    \includegraphics[width=0.8\textwidth]{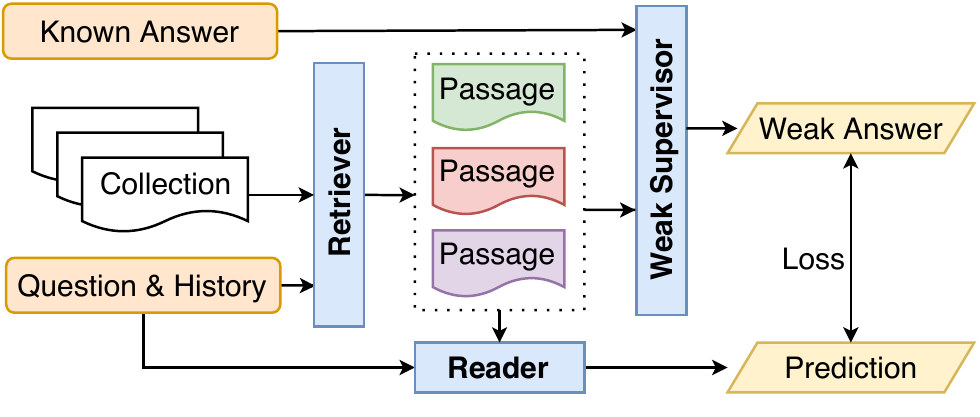}
    \caption{Architecture of our full model. Given a question and its conversation history, the retriever first retrieves top-K relevant passages from the collection. The reader then reads the top passages and produces an answer. We adopt a weakly-supervised training approach. Given the known answer and one of the retrieved passages, the weak supervisor predicts a span in this passage as the weak answer to provide weak supervision signals for training the reader.}
    \label{fig:architecture}
\end{figure} 

We follow the same architecture of the ORConvQA model in Qu et al.~\cite{orconvqa}.\footnote{We disable the reranker in Qu et al.~\cite{orconvqa} since our preliminary experiments indicated the weak supervision signals seem to lead to degradation for reranker and retriever. } 
Our approach differs from theirs in how we train the model. They use full supervision while we adopt weak supervision.
We briefly describe the architecture of this ORConvQA model before introducing our weakly-supervised training approach.

As illustrated in Figure~\ref{fig:architecture}, the ORConvQA model is composed of a passage retriever and a passage reader that are both learnable and based on Transformers~\cite{transformer}. Given a question and its history, the retriever first retrieves top-K relevant passages from the collection. The reader then reads the top passages and produces an answer. History modeling is enabled in both components by concatenating history questions. Since we do not have access to ground-truth history answers and gold passages, advanced history modeling approaches proposed in previous research~\cite{ham,hae} does not apply here. The training contains two phases, a pretraining phase for the retriever, and a concurrent learning phase for the reader and fine-tuning the question encoder in the retriever. Our weakly-supervised training approach is applied to the concurrent learning phase. 

\subsubsection{\textbf{Retriever}}
\label{subsubsec:retriever}
The learnable retriever follows a dual-encoder architecture~\cite{reqa,orqa,Das2019MultistepRI} that has a passage encoder and a question encoder. Both encoders are based on ALBERT~\cite{albert} and can encode a question/passage into a 128-dimensional dense vector. The question is enhanced with history by prepending the initial question and other history questions within a history window. The retriever score is defined as the dot product of the representations of the question and the passage.
The retriever pretraining process ensures the retriever has a reasonable initial performance during concurrent learning. A pretraining example contains a question and its gold passage. Other passages in the batch serve as sampled negatives. 
Using the passage encoder in the pretrained retriever, we encode the collection of passages to a collection of vectors. We then use Faiss\footnote{https://github.com/facebookresearch/faiss}
to create an index of these vectors for maximum inner product search~\cite{mips} on GPU. The question encoder will be fine-tuned during concurrent learning using the retrieved passages. We refer our readers to Qu et al.~\cite{orconvqa} for further details. 

\subsubsection{\textbf{Reader}}
\label{subsubsec:reader}
The reader adapts a standard BERT-based extractive machine comprehension model~\cite{bert} to a multi-document setting by using the shared-normalization mechanism \cite{shared-norm} during training. First, the retrieved passages are encoded independently. Then, the reader maximizes the probabilities of the true start and end tokens among tokens from all the top passages. This step enables the reader to produce comparable token scores across all the retrieved passages for a question. The reader score is defined as the sum of the scores of the start token and the end token. The answer score is then the sum of its retriever score and reader score.

\subsection{Weakly-Supervised Training}
\label{subsec:methods}

The reader component in Qu et al.~\cite{orconvqa} is trained with access to gold passages while our model is supervised by the conversation only. Our weakly-supervised training approach is \textit{more practical} in real-world scenarios. Figure~\ref{fig:architecture} illustrates the role the weak supervisor plays in the system. Given a known answer $a_k$ and one of the retrieved passages $p_j$, the weak supervisor predicts a span in $p_j$ as the \textit{weak answer} $a_k^{weak}$. This weak answer is the weak supervision signal for training the reader. The weak supervisor can also indicate there is no weak answer contained in $p_j$. A question is skipped if there are no weak answers in any of the retrieved passages.

\subsubsection{\textbf{Inspirations}}
\label{subsubsec:inspirations}
Our learned weak supervision method is inspired by the classic span-match weak supervision. This method has been the default and only weak supervision method in previous open-domain QA research~\cite{orqa,drqa,Das2019MultistepRI}. These works mainly focus on factoid QA, where answers are short. A span-match weak supervisor can provide accurate supervision signals since the weak answers are exactly the same as the known answers. In addition, the short answers can find matches easily in passages other than the gold ones.
In information-seeking conversations, however, the answers can be long and freeform, and thus are more difficult to get an exact match in retrieved passages. Although the span-match weak supervisor can still provide accurate supervision signals in this scenario, it renders many training examples useless due to the failure to find exact matches. A straightforward solution is to find a span in a retrieved passage that has the maximum overlap with the known answer. Such overlap can be measured by word-level F1. This overlap method, however, can be intractable and inefficient since it has to enumerate all spans in the passage. 
This method also requires careful tuning for the threshold to output ``no answer''. 
Therefore, we introduce a learned weak supervisor based on Transformers~\cite{transformer} to predict a weak answer span directly in a retrieved passage given the known answer. 
This supervisor also has the ability to indicate that the retrieved passage does not have a good weak answer.

\subsubsection{\textbf{Learned Weak Supervisor}}
\label{subsubsec:supervisor}
\begin{figure}[t]
    \centering
    \includegraphics[width=\textwidth]{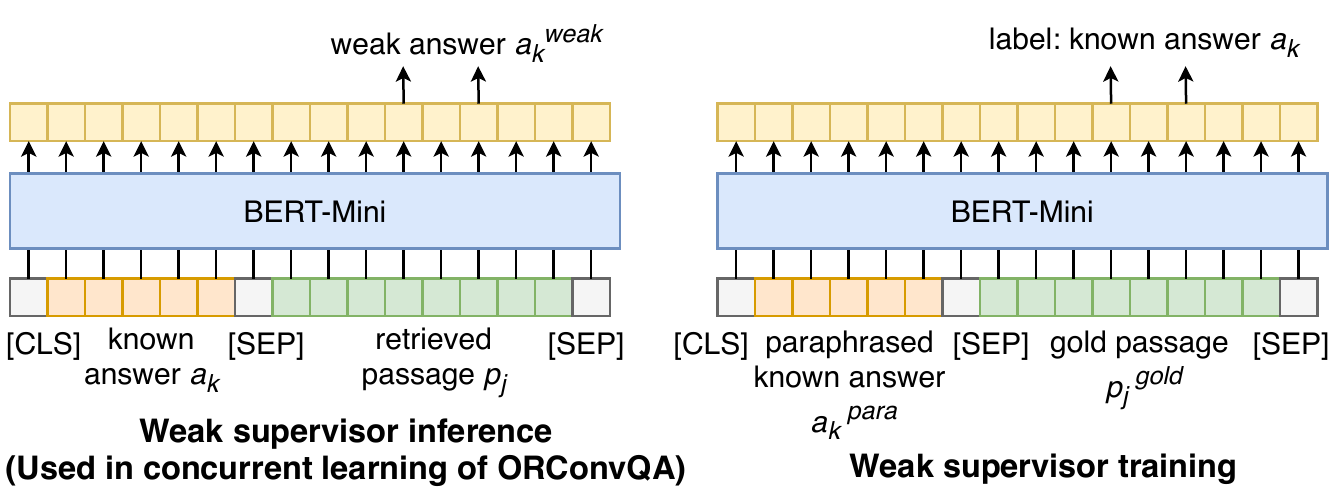}
    \caption{Learned weak supervisor. 
    During the concurrent learning phase of ORConvQA, the weak supervisor conducts inference on a retrieved passage $p_j$ (the left figure) to predict a passage span that is a paraphrase of the known answer $a_k$. 
    When training of the weak supervisor (the right figure), the model is trained to predict the known answer $a_k$ in the passage given a paraphrase of the known answer $a_k^{para}$ and the passage.
    }
    \label{fig:weak-supervisor}
\end{figure} 

Given the known answer $a_k$ and one of the retrieved passages $p_j$, the weak supervisor predicts a span in $p_j$ as the weak answer $a_k^{weak}$. Intuitively, $a_k^{weak}$ is a paraphrase of $a_k$. We use a standard BERT-based extractive MC model~\cite{bert} here as shown in Figure~\ref{fig:weak-supervisor}, except that we use $a_k$ for the question segment. The best weak answer for all top passages is the one with the largest sum of start and end token scores.

Although theoretically simple, this model presents challenges in training because position labels of $a_k^{weak}$ are not available. Therefore, we consider the known answer $a_k$ as the weak answer we are seeking since we know the exact position of $a_k$ in its gold passage $p_j^{gold}$. We then use a diverse paraphrase generation model (described in Section \ref{subsubsec:diverse_paraphraser})
to generate a paraphrase $a_k^{para}$ for the known answer $a_k$. The paraphrase $a_k^{para}$ simulates the known answer during the training of the weak supervisor, as shown in Figure~\ref{fig:weak-supervisor}. The weak supervisor is trained before concurrent learning and kept frozen during concurrent learning. We train the weak supervisor to tell if the passage does not contain a weak answer by pairing a randomly sampled negative passage with the known answer. 

We are aware of a dataset, CoQA~\cite{coqa}, that provides both span answer and freeform answer for a given question $q_k$. In this case, we can take the freeform answer as a natural paraphrase $a_k^{para}$ for the span answer (known answer) $a_k$ when training the weak supervisor. For datasets that do not offer both answer types, our diverse paraphraser assumes the role of the oracle to generate the paraphrase answer. In other words, the use of the diverse paraphraser ensures that our weak supervision approach can be applied to a wide variety of conversation data that are beyond datasets like CoQA.

\subsubsection{\textbf{Diverse Paraphrase Model}}
\label{subsubsec:diverse_paraphraser}
We now briefly describe the diverse paraphraser~\cite{diverse_paraphrase} used in the training process of the learned weak supervisor. This model is built by fine-tuning GPT2-large~\cite{radford2019language} using encoder-free seq2seq modeling~\cite{wolf2018transfertransfo}. 
As training data we use \textsc{paraNMT-50M}~\cite{wieting-gimpel-2018-paranmt}, a massive corpus of back translated data~\cite{wieting-gimpel-2018-paranmt}. The training corpus is aggressively filtered to leave sentence pairs with high lexical and syntactic diversity so that the model can generate diverse paraphrases. We refer our readers to Krishna et al.~\cite{diverse_paraphrase} for further details. 
\section{Experiments}
\label{sec:exp}
We now describe the experimental setup and report the results of our evaluations.
\subsection{Experimental Setup}
\label{subsec:setup}
\subsubsection{\textbf{Dataset}}
\label{subsubsec:data}
We select two ConvQA datasets, QuAC~\cite{quac} and CoQA~\cite{coqa}, with different answer types (span/freeform) to conduct a comprehensive evaluation of our weak supervision approach and to provide insights for weakly-supervised ORConvQA. We present the data statistics of both datasets in Table~\ref{tab:data-stat}. We remove unanswerable questions in both datasets since there is no basis to find weak answers.\footnote{This difference in the data accounts for the discrepancies of the full-supervision results presented in Table~\ref{tab:results-span}.}

\paragraph{OR-QuAC (span answers)}
We use the OR-QuAC dataset introduced in Qu et al.~\cite{orconvqa}. This dataset adapts QuAC to an open-retrieval setting. It contains information-seeking conversations from QuAC, and a collection of 11 million Wikipedia passages (document chunks). 

\paragraph{OR-CoQA (freeform answers)}
We process the CoQA dataset~\cite{coqa} in the Wikipedia domain for the open-retrieval setting following Qu et al.~\cite{orconvqa}, resulting in the OR-CoQA dataset. CoQA offers freeform answers generated by people in addition to span answers, resulting in more natural conversations. OR-CoQA and OR-QuAC share the same passage collection. Similar to QuAC, many initial questions in CoQA are also ambiguous and hard to interpret without the given gold passage (e.g., ``When was the University established?''). OR-QuAC deals with this by replacing the \textit{first question} of a conversation with its context-independent \textit{rewrite} offered by the CANARD dataset~\cite{canard} (e.g., ``When was the University \textit{of Chicago} established?''). This makes the conversations self-contained. Since we are not aware of any CANARD-like resources for CoQA, we prepend the document title to the first question for the same purpose (e.g., ``\textit{University of Chicago} When was the University established?''). 
Since the CoQA test set is not publicly available, we take the original development set as our test set and 100 dialogs from the original training set as our development set.

\begin{table}[t]
\caption{Data Statistics.}
\centering
\label{tab:data-stat}
\begin{tabular}{ l | lll | lll }
\toprule
\multirow{2}{*}{Items}                                                      & \multicolumn{3}{c|}{OR-CoQA} & \multicolumn{3}{c}{OR-QuAC} \\ \cmidrule(l){2-7} 
                                                                            & Train    & Dev     & Test    & Train   & Dev     & Test    \\ \midrule
\# Dialogs                                                                  & 1,521    & 100     & 100     & 4,383   & 490     & 771     \\
\# Questions                                                                & 23,027   & 1,494   & 1,611   & 25,824  & 2,808   & 4,406   \\
\# Avg. Question Tokens                                                     & 5.8      & 5.7     & 5.8     & 6.8     & 6.6     & 6.8     \\
\# Avg. Answer Tokens                                                       & 2.8      & 2.6     & 2.6     & 15.0    & 15.0    & 14.7    \\
\# Avg. Dialog Questions                                                    & 15.1     & 14.9    & 16.1    & 5.9     & 5.7     & 5.7     \\
\begin{tabular}[c]{l} \# Avg./Max History \\ Turns per Question\end{tabular} & 7.9/22   & 7.6/21  & 7.9/19  & 2.8/11  & 2.8/11  & 2.8/11  \\ \bottomrule
\end{tabular}
\end{table}

\subsubsection{\textbf{Competing Methods}}
\label{subsubsec:competing-methods}
Since this work focuses on weak supervision, we use the same ORConvQA model and vary the supervision methods. To be specific, the competing methods are:
\begin{itemize}
    \item \textbf{Full supervision} (Full S): Manually add the gold passage to the retrieval results and use the ground-truth answer span \cite{orconvqa}. This only applies to QuAC since we have no passage relevance for CoQA. This method serves as the upper bound of model performance and it is not comparable with other weak supervision methods that do not have access to the groundtruth answers in concurrent learning.
    
    \item \textbf{Span-match weak supervision} (Span-match WS): This method finds a weak answer span that is identical to the known answer in the retrieved passages. When there are multiple matched spans, we take the first one.
    
    \item \textbf{Learned weak supervision} (Learned WS): This is our method in Section~\ref{subsec:methods} that finds a paraphrased span of the known answer as the weak answer.
    
    \item \textbf{Combined weak supervision} (Combined WS): 
    This is the combination of the above two methods. 
    We first use the span-match weak supervisor to try to find a weak answer. If it fails, we take the weak answer found by the learned weak supervisor.
\end{itemize}

\subsubsection{\textbf{Evaluation Metrics}}
\label{subsubsec:metrics}
We use the word-level F1 and human equivalence score (HEQ) \cite{quac} to evaluate the performance of ConvQA. \textbf{F1} evaluates the overlap between the prediction and the ground-truth answer. \textbf{HEQ} is the percentage of examples for which system F1 $\geq$ human F1. This is computed on a question level (HEQ-Q) and a dialog level (HEQ-D).

In addition to the performance metrics described above, we define another set of metrics to reveal the impact of the weak supervisor in the training process as follows. 
\textbf{\% Has Answer} is the percentage of training examples that have a weak answer (in the last epoch). 
\textbf{\% Hit Gold} is the percentage of training examples that have a weak answer identified in gold passages (in the last epoch). 
\textbf{Recall} is the percentage of training examples that have the gold passage retrieved  (in the last epoch). 
\textbf{\% From Gold} is the percentage of predicted answers that are extracted from the gold passages.

\subsubsection{\textbf{Implementation Details}}
\label{subsubsec:details}
Our models are based on the open-source implementation of ORConvQA\footnote{\url{https://github.com/prdwb/orconvqa-release}}, Diverse Paraphrase Model\footnote{\url{https://github.com/martiansideofthemoon/style-transfer-paraphrase}}, and the HuggingFace Transformers repository.\footnote{https://github.com/huggingface/transformers} We use the same pretrained retriever in Qu et al.~\cite{orconvqa} for both datasets.
For concurrent learning of ORConvQA, we set the number of training epochs to 5 (larger than \cite{orconvqa}) to account for the skipped steps where no weak answers are found. We set the number of passages to update the retriever to 100, and the history window size to 6 since these are the best settings reported in \cite{orconvqa}. The max answer length is set to 40 for QuAC and 8 for CoQA. The rest of the hyper-parameters and implementation details for the ORConvQA model are the same as in \cite{orconvqa}. 

For the weak supervisor, we use BERT-Mini~\cite{smaller-bert} for better efficiency.
We set the number of training epochs to 4, the learning rate to 1e-4, and the batch size to 16. 
As discussed in Section~\ref{subsubsec:supervisor}, the diverse paraphraser is used for OR-QuAC only. For OR-CoQA, we use the freeform answer provided by the dataset as a natural paraphrase to the span answer.

\subsection{Evaluation Results on Span Answers}
\label{subsec:results-span}
\setlength{\tabcolsep}{5pt}
\begin{table}[t]
\caption{Evaluation results on OR-QuAC (span answers). The learned weak supervisor causes no statistical significant performance decrease compared span match.}
\label{tab:results-span}
\centering
\begin{tabular}{@{}c|c|cccc@{}}
\toprule
\multicolumn{2}{c|}{Methods}          & Full S        & Span-match WS & Learned  WS  & Combined WS   \\ \midrule
Train                 & \% Has Answer & 100.00\%      & 72.96\%       & 75.98\%      & 75.52\%       \\ \midrule
\multirow{3}{*}{Dev}  & F1            & \textbf{22.8} & \textbf{20.8} & 20.2         & 20.1          \\
                      & HEQ-Q         & \textbf{8.1}  & \textbf{6.8}  & 6.0          & 6.4           \\
                      & HEQ-D         & 0.6           & 0.6           & 0.2          & 0.6           \\ \midrule
\multirow{3}{*}{Test} & F1            & \textbf{23.9} & \textbf{23.6} & 23.1         & 23.2          \\
                      & HEQ-Q         & \textbf{14.0} & 12.3          & 11.8         & \textbf{12.5} \\
                      & HEQ-D         & \textbf{2.2}  & 1.7           & \textbf{1.9} & \textbf{1.9}  \\ \bottomrule
\end{tabular}
\end{table}

Given the different properties of span answers and freeform answers, we study the performance of our weak supervision approach on these answers separately.
We report the evaluation results on the span answers in Table~\ref{tab:results-span}. Our observations can be summarized as follows.

The full supervision setting yields the best performance, as expected. This verifies the supervision signals provided by the gold passages and the ground-truth answer spans are more accurate than the weak ones. 
Besides, all supervision approaches have similar performance on span answers. This suggests that span-match weak supervision is sufficient to handle conversations with span answers. Ideally, if the known answer is part of the given passage, the learned weak supervisor should be able to predict the weak answer as exactly the same with the known answer. In other words, the learned weak supervisor should fall back to the span-match weak supervisor when handling span answers. In practice, this is not guaranteed due to the variance of neural models. However, our learned weak supervisor causes no statistical significant performance decrease compared with the span-match supervisor. This demonstrates that the learned weak supervision approach can cover span answers as well.
Although we observe that the learned supervisor can identify more weak answers than span match, these weak answers could be false positives that do not contribute to the model performance.
Finally, for the combined weak supervisor, our analysis shows that 96\% of the weak answers are identified by span match, further explaining the fact that all weak supervision approaches have almost identical performance.

\subsection{Evaluation Results on Freeform Answers}
\label{subsec:results-freeform}
We then look at the evaluation results on freeform answers in Table~\ref{tab:results-freeform}. These are the cases where a span-match weak supervisor could fail. We observe that combining the learned weak supervisor with span match brings a statistically significant improvement over the span-match baseline on the test set, indicating these two methods complement each other. The test set has multiple reference answers per question, making the evaluation more practical.
In addition, the learned supervisors can identify more weak answers than span match, these weak answers contribute to the better performance of our model.
Further, for the combined weak supervisor, our analysis shows that 77\% of the weak answers are identified by span match. This means that nearly a quarter of the weak answers are provided by the learned supervisor and used to improve the performance upon span match. This further validates the source of effectiveness of our model.

\setlength{\tabcolsep}{5pt}
\begin{table}[t]
\caption{Evaluation results on OR-CoQA (freeform answers). $\ddagger$ means statistically significant improvement over the span-match baseline with $p < 0.05$. 
}
\label{tab:results-freeform}
\centering
\begin{tabular}{@{}c|c|ccc@{}}
\toprule
\multicolumn{2}{c|}{Methods}          & Span-match WS & Learned  WS & Combined WS              \\ \midrule
Train                 & \% Has answer & 51.81\%       & 65.75\%     & 70.35\%                  \\ \midrule
\multirow{3}{*}{Dev}  & F1            & 18.3          & 18.9        & \textbf{19.7}            \\
                      & HEQ-Q         & 11.6          & 9.0         & \textbf{12.7}            \\
                      & HEQ-D         & 0.0           & 0.0         & 0.0                      \\ \midrule
\multirow{3}{*}{Test} & F1            & 24.3          & 26.0        & \textbf{28.8}$^\ddagger$ \\
                      & HEQ-Q         & 19.9          & 15.9        & \textbf{22.5}            \\
                      & HEQ-D         & 0.0           & 0.0         & 0.0                      \\ \bottomrule
\end{tabular}
\end{table}

\subsection{A Closer Look at the Training Process}
\label{subsec:closer-look}
\begin{table}[t]
\caption{A closer look at the training process for OR-QuAC. }
\centering
\label{tab:closer-look}
\begin{tabular}{@{}c|ccc|c|c@{}}
\toprule
\multirow{2}{*}{Methods} & \multicolumn{3}{c|}{Train}        & \multicolumn{1}{c|}{Dev} & \multicolumn{1}{c}{Test} \\ \cmidrule(l){2-6} 
                        & \% Has Ans & \% Hit Gold & Recall & \% From Gold             & \% From Gold             \\ \midrule
Full S            & 100.00\%   & 100.00\%    & 1.0000 & 45.23\%                  & 27.46\%                  \\
Span-match WS            & 72.96\%    & 68.97\%     & 0.7190 & 40.88\%                  & 28.80\%                  \\
Learned WS                 & 75.98\%    & 67.24\%     & 0.7187 & 39.89\%                  & 28.73\%                  \\
Combined WS           & 75.52\%    & 68.37\%     & 0.7129 & 40.28\%                  & 28.39\%                  \\ \bottomrule
\end{tabular}
\end{table}

We take a closer look at the training process, as shown in Table~\ref{tab:closer-look}. We conduct this analysis on OR-QuAC only since we do not have the ground-truth passage relevance for CoQA. We observe that, ``\% Has Ans'' are higher than ``\% Hit Gold'' for all weak supervision methods, indicating all of them can identify weak answers in passages beyond the gold passages. In particular, our method can identify more weak answers than span match. 
We also notice that ``\% Hit Gold'' is only slightly lower than ``Recall'', suggesting that most of the retrieved gold passages can yield a weak answer. This verifies the capability of weak supervisors.
Finally, ``\% From Gold'' are relatively low for all methods, indicating great potential for improvements.

\subsection{Case Study and Error Analysis}
\label{subsec:case-study}

\begin{table*}[t]
\caption{Case study. Weak answers are found by the learned weak supervisor. Boldface denotes discrepancies and italic denotes paraphrasing.}
\centering
\label{tab:case-study}
\begin{tabular}{@{}c|c|ll@{}}
\toprule
              & \#                 & \multicolumn{2}{c}{Questions and Answers}                                                                                                                                                                                     \\ \midrule
\multirow{9}{*}{\rotatebox[origin=c]{90}{Good}} & \multirow{3}{*}{1} & \multicolumn{1}{l|}{Question}     & Where was the album released?                                                                                                                                                             \\
                      &                    & \multicolumn{1}{l|}{Known answer} & on online forums and music sites.                                                                                                                                                         \\
                      &                    & \multicolumn{1}{l|}{Weak answer}  & on online forums and music sites.                                                                                                                                                         \\ \cmidrule(l){2-4} 
                      & \multirow{3}{*}{2} & \multicolumn{1}{l|}{Question}     & ... mention anything else he starred in?                                                                                                                                                  \\
                      &                    & \multicolumn{1}{l|}{Known answer} & After starring ... the film adaptation of The Music Man                                                                                                                                   \\
                      &                    & \multicolumn{1}{l|}{Weak answer}  & After starring ... film adaptation of The Music Man \textbf{(1962)}.                                                                                                                      \\ \cmidrule(l){2-4} 
                      & \multirow{3}{*}{3} & \multicolumn{1}{l|}{Question}     & Where did he distribute the Cocaine?                                                                                                                                                      \\
                      &                    & \multicolumn{1}{l|}{Known answer} & \begin{tabular}[c]{@{}l@{}}flying out planes several times, mainly between Colombia and \\ Panama, along smuggling routes into the United States.\end{tabular}                           \\
                      &                    & \multicolumn{1}{l|}{Weak answer}  & \begin{tabular}[c]{@{}l@{}}\textit{He flew a plane himself several times, mainly between Colombia}\\ \textit{and Panama, in order to smuggle a load into the United States.}\end{tabular} \\ \midrule
\multirow{6}{*}{\rotatebox[origin=c]{90}{Bad}}  & \multirow{3}{*}{4} & \multicolumn{1}{l|}{Question}     & how long have people had clothes?                                                                                                                                                         \\
                      &                    & \multicolumn{1}{l|}{Known answer} & as long ago as 650 thousand years ago                                                                                                                                                     \\
                      &                    & \multicolumn{1}{l|}{Weak answer}  & around \textbf{170,000} years ago.                                                                                                                                                        \\ \cmidrule(l){2-4} 
                      & \multirow{3}{*}{5} & \multicolumn{1}{l|}{Question}     & What is data compression called?                                                                                                                                                          \\
                      &                    & \multicolumn{1}{l|}{Known answer} & reducing the size of a data file                                                                                                                                                          \\
                      &                    & \multicolumn{1}{l|}{Weak answer}  & \textbf{By using wavelets, a compression ratio}                                                                                                                                           \\ \bottomrule
\end{tabular}
\end{table*}

We then conduct a qualitative analysis by presenting weak answers identified by the learned weak supervisor in Table~\ref{tab:case-study} to better understand the weak supervision process. Example 1 and 2 show that our learned weak supervisor can find weak answers that are exactly the same or almost identical to the known answers when an exact match of the known answer exits, further validating our method can potentially cover span-match weak supervision. Example 3 shows that if an exact match does not exist, our method can find a weak answer that expresses the same meaning with the known answer. This is a case that a span-match weak supervisor would fail.

Example 4 shows that our method tends to focus on the lexical similarity only but get the fact wrong. Example 5 indicates our method sometimes finds a weak answer that is relevant to the known answer but cannot be considered as a good answer. These are the limitations of our method.
\section{Conclusions and Future Work} 
\label{sec:conclusion}

In this work, we propose a learned weak supervision approach for open-retrieval conversational question answering. Extensive experiments on two datasets show that, although span-match weak supervision can handle span answers, it is not sufficient for freeform answers. Our learned weak supervisor is more flexible since it can handle both span answers and freeform answers. It is more powerful when combined with the span-match supervisor. For future work, we would like to enhance the performance of ORConvQA by studying more advanced history modeling methods and more effective weak supervision approaches.

\subsubsection{\textbf{Acknowledgments}}
This work was supported in part by the Center for Intelligent Information Retrieval and in part by NSF IIS-1715095. Any opinions, findings and conclusions or recommendations expressed in this material are those of the authors and do not necessarily reflect those of the sponsor. The authors would like to thank Minghui Qiu for his constructive comments on this work.

\bibliographystyle{splncs04}
\bibliography{acmart}

\end{document}